\documentclass[twocolumn,showpacs,preprintnumbers]{revtex4}
\usepackage{amsmath}
\usepackage{graphicx}
\usepackage{dcolumn}
\usepackage{bm}
\usepackage{color}

\begin{document}

\title{$\mathcal{PT}$ symmetry of a square-wave modulated  two-level system}

\author{Liwei Duan$^{1}$}
\author{Yan-Zhi Wang$^{1}$}
\author{Qing-Hu Chen$^{1,2}$}\email{qhchen@zju.edu.cn}

\affiliation{
$^{1}$ Department of Physics and Zhejiang Province Key Laboratory of Quantum Technology and Device, Zhejiang University, Hangzhou 310027, China \\
$^{2}$ Collaborative Innovation Center of Advanced Microstructures,  Nanjing University,  Nanjing 210093, China}
\date{\today }

\begin{abstract}
We study a non-Hermitian two-level system with square-wave modulated dissipation and coupling. Based on the Floquet theory, we achieve an effective Hamiltonian from which the boundaries of the $\mathcal{PT}$ phase diagram are captured exactly. Two kinds of $\mathcal{PT}$ symmetry broken phases are found whose effective Hamiltonians differ by a constant $\omega / 2$. For the time-periodic dissipation, a vanishingly small dissipation strength can lead to the $\mathcal{PT}$ symmetry  breaking in the $(2k-1)$-photon resonance  ($\Delta = (2k-1) \omega$),  with $k=1,2,3\dots$ It is worth noting that such a phenomenon can also happen in $2k$-photon resonance  ($\Delta = 2k \omega$), as long as the dissipation strengths or the driving times are imbalanced, namely $\gamma_0 \ne - \gamma_1$ or $T_0 \ne T_1$. For the time-periodic coupling, the weak dissipation induced $\mathcal{PT}$ symmetry breaking occurs at $\Delta_{\mathrm{eff}}=k\omega$, where $\Delta_{\mathrm{eff}}=\left(\Delta_0 T_0 + \Delta_1 T_1\right)/T$. In the high frequency limit, the phase boundary is given by a simple relation $\gamma_{\mathrm{eff}}=\pm\Delta_{\mathrm{eff}}$.
\end{abstract}


\maketitle

\section{Introduction}

A non-Hermitian Hamiltonian is a natural extension of the conventional Hermitian one to describe the open quantum system.
The discovery of the real spectra in non-Hermitian Hamiltonians by Bender and Boettcher \cite{PhysRevLett.80.5243} has stimulated enormous interests in the systems with parity-time  ($\mathcal{PT}$) symmetry \cite{PhysRevLett.89.270401,doi:10.1080/00107500072632,
Bender_2007,RevModPhys.88.035002}.
Early theoretical and  experimental explorations of the non-Hermitian systems with $\mathcal{PT}$ symmetry mainly focus on the optics and photonics \cite{Ruschhaupt_2005,El-Ganainy:07,PhysRevLett.100.103904,
PhysRevLett.100.030402,PhysRevLett.103.093902,ruter2010observation,
Feng729,regensburger2012parity,Hodaei975,PhysRevLett.110.243902,
weimann2017topologically,PhysRevLett.123.230401}. Feng \textit{et al.} realized the nonreciprocal light propagation in a Silicon photonic circuit which provides a way to chip-scale optical isolators for optical communications and computing \cite{Feng729}. Hodaei \textit{et al.}  stabilized single-longitudinal mode operation in a system of coupled mirroring lasers by harnessing notions from $\mathcal{PT}$ symmetry, which provides the possibilities to develop optical devices with enhanced functionality \cite{Hodaei975}. Xiao \textit{et al.} achieved the first experimental characterization of critical phenomena in $\mathcal{PT}$-symmetric nonunitary quantum dynamics \cite{PhysRevLett.123.230401}. Recent experiments have realized the non-Hermitian Magnon-polaritons systems, and higher-order exceptional points were observed which can be used to measuring the output spectrum of the cavity \cite{zhang2017observation,PhysRevB.99.054404,shen68research}. The anomalous edge state in a non-Hermitian lattice \cite{PhysRevLett.116.133903} has intrigued  persistent attention to the combination of the non-Hermiticity and the topological phase \cite{PhysRevLett.120.146402,PhysRevLett.121.026808,PhysRevX.8.031079,Kawabata2018Topological,PhysRevX.9.041015,
PhysRevLett.118.040401,PhysRevLett.122.076801,PhysRevLett.121.086803,PhysRevLett.123.170401,
xiao2017observation,PhysRevLett.121.136802,PhysRevLett.123.066404,J_S_Liu:10302,J_S_Liu:100304}.
The non-Bloch band theory has been developed to describe the non-Hermitian lattice systems \cite{PhysRevLett.120.146402,PhysRevLett.121.086803,PhysRevLett.121.136802,PhysRevLett.123.066404}. Kawabata \textit{et al.} established a fundamental symmetry principle in non-Hermitian physics which paved the way towards a unified framework for non-equilibrium topological phase \cite{Kawabata2018Topological,PhysRevX.9.041015}. Yao \textit{et al.} studied the bulk-boundary correspondence in the non-Hermitian systems and found the non-Hermitian skin effect \cite{PhysRevLett.121.086803,PhysRevLett.123.170401}. Xiao \textit{et al.} observed the topological edge states in $\mathcal{PT}$-symmetric quantum walks \cite{xiao2017observation}.

Recently, Joglekar \textit{et al.} investigated a two-level system coupled to a sinusoidally varying gain-loss potential, namely, the non-Hermitian Rabi model with time-periodic dissipation \cite{PhysRevA.90.040101}. They found that there existed multiple frequency windows where $\mathcal{PT}$ symmetry was broken and restored. The non-Hermitian Rabi model has drawn growing attention due to its especially rich phenomena which are absent in the static counterparts \cite{PhysRevA.92.042103,PhysRevA.91.042134,PhysRevA.83.052125,PhysRevA.91.042135,
PhysRevA.99.012107,PhysRevA.98.052122,PhysRevA.95.052128,PhysRevLett.119.093901,li2019observation,Chun-Yu_Jia:40502}. Lee \textit{et al.} found the $\mathcal{PT}$ symmetry breaking  at the $(2k-1)$-photon resonance and derived the boundaries of the $\mathcal{PT}$ phase diagram by doing perturbation theory beyond rotating-wave approximation \cite{PhysRevA.92.042103}. Gong \textit{et al.} found that a periodic driving could stabilize the dynamics despite the loss and gain in the non-Hermitian system \cite{PhysRevA.91.042135,PhysRevA.99.012107}. Xie \textit{et al.} studied a non-Hermitian Rabi model with time-periodic coupling and found exact analytical results for certain exceptional points \cite{PhysRevA.98.052122}. A synchronous modulation which combined the time-periodic dissipation and coupling was study  in Ref. \cite{PhysRevA.95.052128}, which provided an additional possibility for pulse manipulation and coherent control of the $\mathcal{PT}$-symmetric two-level systems. Experimental approach of Floquet $\mathcal{PT}$-symmetric system has been proposed with two coupled high frequency oscillators \cite{PhysRevLett.119.093901}. A $\mathcal{PT}$ symmetry breaking transition by engineering time-periodic dissipation and coupling has been realized through state-dependent atom loss in an optical dipole trap of ultracold $^6\mathrm{Li}$ atoms \cite{li2019observation}. They confirmed that a weak time-periodic dissipation could lead to $\mathcal{PT}$-symmetry breaking in $(2k-1)$-photon resonance.  {It should be noted that the $\mathcal{PT}$-symmetry breaking can occur in a finite non-Hermitian system, which is quite different from the quantum phase transition in the Hermitian system where the thermodynamic limit is needed \cite{li2019observation,LEE2014}.}

In this paper, we study the $\mathcal{PT}$ symmetry of a  two-level system with time-periodic dissipation and coupling. Instead of the widely used sinusoidal modulation\cite{PhysRevA.90.040101,PhysRevA.92.042103,
PhysRevA.91.042135,PhysRevA.99.012107,PhysRevA.98.052122}, we consider a  square-wave one, which is easier to implement in the ultracold atoms experiment \cite{li2019observation} and has analytical exact solutions based on the Floquet theory \cite{PhysRev.138.B979,PhysRevA.7.2203}. The square-wave modulation has a broad range of applications in the Hermitian system. It has been used to suppress the quantum dissipation in spin chains \cite{PhysRevA.91.052122}, to generate many Majorana modes in  a one-dimensional p-wave superconductor system \cite{PhysRevB.87.201109}, to generate large-Chern-number topological phases \cite{PhysRevB.93.184306}, and so on. The square-wave modulation has also been realized in the non-Hermitian systems \cite{li2019observation}. This paper is organized as follows. In section II, we describe the non-Hermitian Hamiltonian of the driving two-level system. In section III, we briefly introduce the Floquet theory and derive the effective static Hamiltonian. In section IV, we achieve the $\mathcal{PT}$ phase diagram and analyze the influence of multiphoton resonance. An equivalent Hamiltonian is obtained in the high frequency limit.  The last section contains some concluding remarks.

\section{Hamiltonian}

We consider a periodically driving two-level system $H(t) = H(t + T)$, with
\begin{eqnarray}
H\left( t\right) =\frac{\Delta\left(
t\right) }{2}\sigma _{x}+\mathrm{i} \frac{\gamma \left(
t\right) }{2}\sigma _{z}, \label{Ht}
\end{eqnarray}%
where $\sigma_{x, z}$ are the Pauli matrices, $T = T_0 + T_1$ is the driving period, $\omega =2\pi /T$ is the
driving frequency, $\Delta(t)$ is the time-periodic coupling strength, and $\gamma(t)$ is the dissipation strength which leads to the
periodic gain and loss. Lee \textit{et al.} \cite{PhysRevA.92.042103} studied
the $\mathcal{PT}$ phase diagram of the non-Hermitian two-level system by doing
the perturbation theory, which corresponds to $\Delta(t)=\Delta$ and $\gamma (t) = 4 \lambda
\cos(\omega t)$. Xie \textit{et al.} \cite{PhysRevA.98.052122} found the exact analytical results for certain exceptional points  of the two-level system with time-periodic coupling, which corresponds to $\Delta(t)=v_0 + v_1 \cos(\omega t)$ and $\gamma (t) = \gamma$. Luo \textit{et al.} \cite{PhysRevA.95.052128} studied the analytical results of the non-Hermitian two-level systems with sinusoidal modulations of both $\Delta(t)$ and $\gamma(t)$.  In order to get the exact analytical results without  {using
perturbation theory}, we consider a synchronous square-wave modulation of both dissipation and coupling. The corresponding time-periodic parameters are
\begin{equation}
f \left( t\right) =\left\{%
\begin{array}{ll}
f _{0} & \mathrm{, ~if }~~ m T - \frac{T_0}{2} \leq t < m T + \frac{T_0}{2}, \\
f _{1} & \mathrm{, ~if }~~ m T + \frac{T_0}{2} \leq t < (m + 1)T - \frac{T_0}{2},%
\end{array}%
\right.
\end{equation}
with $f=\Delta$, $\gamma$ and $m=\dots, -1,0,1, \dots$
It's easy to confirm that the non-Hermitian Hamiltonian has a $\mathcal{PT}$ symmetry,  {namely $\hat{\mathcal{P}} H^{\dagger}(t)  \hat{\mathcal{P}} = H(t)$, where $H^{\dagger}(t)$ is the Hermitian conjugate of $H(t)$ and { $\hat{\mathcal{P}} = \hat{\mathcal{P}}^{-1}  =  \sigma_x $} is the parity operator \cite{PhysRevLett.89.270401,RevModPhys.88.035002}}. This non-Hermitian system has been realized by Li \textit{et al.} in the ultracold atoms experiments \cite{li2019observation}. However, they focused on a special case with only one time-periodic parameter (either dissipation or coupling), and  $f_{0}=f$, $f_{1}=0$, $T_0=T_1=T/2$. We consider a more general case which relieves those constraints.
Two time-independent Hamiltonians $H_0$ and $H_1$ appear
alternately, with
\begin{eqnarray}
H_j =\frac{\Delta_j }{2}\sigma _{x}+\mathrm{i}\frac{\gamma_j }{2}\sigma _{z}, ~~
j=0, 1,
\end{eqnarray}
and the corresponding eigenenergies are $E_j^{\pm} = \pm h_j$ where
\begin{eqnarray}
h_{j} = \frac{\sqrt{ \Delta_j ^{2}-\gamma _{j}^{2}}}{2}.
\end{eqnarray}
$H_j$ is one of the simplest non-Hermitian systems with $\mathcal{PT}$
symmetry \cite{RevModPhys.88.035002}. When $|\Delta_j| > |\gamma_j|$, the eigenenergy is real and it
corresponds to the $\mathcal{PT}$-symmetric phase. When $|\Delta_j| < |\gamma_j|$%
, the eigenenergy is imaginary and the $\mathcal{PT}$
symmetry is  broken. When $|\Delta_j| = |\gamma_j|$, there exists an
exceptional point (EP). The dynamics at each time domain is governed by the
time evolution operator
\begin{eqnarray}
U_j(T_{j}) &=& \exp \left( -iH_{j}T_{j}\right) \nonumber\\
&=&\cos \left( h_{j}T_{j}\right)
I -\mathrm{i} \mathrm{sinc} \left( h_{j}T_{j}\right) T_{j} H_{j},  \label{Uj}
\end{eqnarray}
where $I$ is a $2 \times 2$ identity matrix, and $\mathrm{sinc}(x)=\sin(x)/x$.

\section{Floquet theory}

According to the Floquet theory \cite{PhysRev.138.B979,PhysRevA.7.2203}, we can define an effective Hamiltonian $H_{%
\mathrm{eff}}$ which satisfies the condition,
\begin{equation}
U_{\mathrm{eff}}(T)=\exp \left( -\mathrm{i} H_{\mathrm{eff}}T\right) =\mathcal{T}\exp %
\left[ -\mathrm{i}\int_{-\frac{T_0}{2}}^{T-\frac{T_0}{2}}dtH(t)\right] .
\end{equation}%
The eigenenergies of the effective Hamiltonian correspond to the Floquet
quasi-energies. Due to the simplicity of the square-wave modulation, the time
evolution operator in a period can be written as
\begin{equation}
\mathcal{T}\exp \left[ -\mathrm{i}\int_{-\frac{T_0}{2}}^{T-\frac{T_0}{2}}dtH(t)\right] =\exp \left(
-\mathrm{i} H_{1}T_{1}\right) \exp \left( -\mathrm{i} H_{0}T_{0}\right) ,
\end{equation}%
Therefore,
\begin{equation}
U_{\mathrm{eff}}(T)=U_{1}(T_{1})U_{0}(T_{0}).\label{Ueff}
\end{equation}%
From Eq. (\ref{Uj}) and (\ref{Ueff}), we achieve the effective time
evolution operator
\begin{widetext}
\begin{eqnarray}
U_{\mathrm{eff}}(T) &=&\left( \cos \left( h_{1}T_{1}\right) \cos \left(
h_{0}T_{0}\right) +\frac{1}{4}\left( \gamma _{1}\gamma _{0}-\Delta_{1}\Delta_{0}\right) T_{1}T_{0}\mathrm{sinc}\left( h_{1}T_{1}\right) \mathrm{sinc}%
\left( h_{0}T_{0}\right) \right) I  \notag \\
&&-i\frac{1 }{2}\left(\Delta_{0} T_{0}\cos \left( h_{1}T_{1}\right) \mathrm{%
sinc}\left( h_{0}T_{0}\right) + \Delta_{1} T_{1}\cos \left( h_{0}T_{0}\right) \mathrm{%
sinc}\left( h_{1}T_{1}\right) \right) \sigma _{x}  \notag \\
&&+\frac{1}{4}  \left( \Delta_{0} \gamma _{1}- \Delta_{1}\gamma _{0}\right) T_{1}T_{0}%
\mathrm{sinc}\left( h_{1}T_{1}\right) \mathrm{sinc}\left( h_{0}T_{0}\right)
\sigma _{y}  \notag \\
&&+\frac{1}{2}\left( \gamma _{0}T_{0}\cos \left( h_{1}T_{1}\right) \mathrm{%
sinc}\left( h_{0}T_{0}\right) +\gamma _{1}T_{1}\cos \left( h_{0}T_{0}\right)
\mathrm{sinc}\left( h_{1}T_{1}\right) \right) \sigma _{z}.\label{Ueff0}
\end{eqnarray}%
\end{widetext}
Since $h_{j}$ can be either a pure real number in the $\mathcal{PT}$%
-symmetric phase or a pure imaginary one in the $\mathcal{PT}$ symmetry
broken phase, both $\cos \left( h_{j}T_{j}\right) $ and $\mathrm{sinc}\left(
h_{j}T_{j}\right) $ must be real numbers. Accordingly, the coefficients
before $I$, $\sigma _{y}$ and $\sigma _{z}$ must be real, while those before
$\sigma _{x}$ must be imaginary. It's easy to confirm that the effective
Hamiltonian can only be the following form,
\begin{eqnarray}
H_{\mathrm{eff}} =\frac{J }{2}\sigma _{x}+\mathrm{i}\left( \frac{\Gamma _{y}}{2}%
\sigma _{y}+\frac{\Gamma _{z}}{2}\sigma _{z}\right) +\frac{n\omega }{2}I, \label{Heff}
\end{eqnarray}%
with $n=0,1$. The eigenenergies of $H_{\mathrm{eff}}$, or the Floquet quasi-energies of $H(t)$ would be $E^{\pm}=\pm h + \frac{n \omega}{2}$,
where
\begin{eqnarray}
h =\frac{\sqrt{ J^{2}-\Gamma_y ^{2}-\Gamma_z ^{2}}}{2}.
\end{eqnarray}
The effective time evolution operator can be rewritten as
\begin{eqnarray}
U_{\mathrm{eff}}(T) &=& \exp \left( -iH_{\mathrm{eff}}T\right) \label{Ueff1} \\
&=&(-1)^n  \cos \left( hT\right) I \nonumber\\
&&- i  \frac{(-1)^n T}{2}\mathrm{sinc}\left( hT\right) J
\sigma _{x}  \nonumber\\
&&+  \frac{(-1)^n T}{2}\mathrm{sinc}\left( hT\right) (\Gamma _{y}\sigma _{y} +\Gamma _{z}\sigma _{z})\nonumber
\end{eqnarray}
By comparing the coefficients before $I$, $\sigma_x$, $\sigma_y$, and $\sigma_z$ in Eq. (\ref{Ueff0}) and Eq. (\ref{Ueff1}), we can directly obtain that
\begin{widetext}
\begin{eqnarray}
(-1)^n \cos\left(h T\right) &=&  \cos \left( h_{1}T_{1}\right) \cos \left(
h_{0}T_{0}\right) +\frac{1 }{4}\left( \gamma _{1}\gamma _{0}-\Delta_{1}
\Delta_{0}\right) T_{0}T_{1}\mathrm{sinc}\left( h_{1}T_{1}\right) \mathrm{sinc}%
\left( h_{0}T_{0}\right) ,\label{cos_h} \\
J &=&  \frac{(-1)^n  }{T\mathrm{sinc}\left( hT\right)}\left(\Delta_{0} T_{0}\cos \left( h_{1}T_{1}\right) \mathrm{%
sinc}\left( h_{0}T_{0}\right) + \Delta_{1} T_{1}\cos \left( h_{0}T_{0}\right) \mathrm{%
sinc}\left( h_{1}T_{1}\right) \right), \label{J}\\
\Gamma_y &=& \frac{(-1)^n}{2T \mathrm{sinc}\left( hT\right)} \left(\Delta_{0} \gamma _{1}- \Delta_{1} \gamma _{0}\right) T_{0}T_{1}%
\mathrm{sinc}\left( h_{1}T_{1}\right) \mathrm{sinc}\left( h_{0}T_{0}\right),\label{Gamma_y}\\
\Gamma_z &=& \frac{(-1)^n}{T \mathrm{sinc}\left( hT\right)} \left( \gamma _{0}T_{0}\cos \left( h_{1}T_{1}\right) \mathrm{%
sinc}\left( h_{0}T_{0}\right) +\gamma _{1}T_{1}\cos \left( h_{0}T_{0}\right)
\mathrm{sinc}\left( h_{1}T_{1}\right) \right). \label{Gamma_z}
\end{eqnarray}
\end{widetext}
Once we get $J$, $\Gamma_y$, $\Gamma_z$ and $n$, the effective Hamiltonian (\ref{Heff}) is finally determined.

\section{Results and Discussions}

The major differences of the effective Hamiltonian and the original one are the dissipation $\Gamma_y$ in $y$-axis and the additional constant $\omega / 2$.
% {We conjecture that the periodic modulation leads to a phase difference which introduces $\Gamma_y$.}
We will show later that the additional constant is closely related with the $\mathcal{PT}$ symmetry broken phases and the exceptional points.
One can easily confirm that the effective Hamiltonian has a $\mathcal{PT}$ symmetry,  {namely $\hat{\mathcal{P}} H^{\dagger}_{\mathrm{eff}} \hat{\mathcal{P}} = H_{\mathrm{eff}}$, since $\hat{\mathcal{P}} \sigma_x \hat{\mathcal{P}} = \sigma_x$, $\hat{\mathcal{P}} \sigma_y \hat{\mathcal{P}} = -\sigma_y$ and $\hat{\mathcal{P}} \sigma_z \hat{\mathcal{P}} = -\sigma_z$}. When  $|\cos\left(h T\right)|<1$, $h$ must be a real number and the $\mathcal{PT}$ symmetry is preserved. For the $\mathcal{PT}$-symmetric phase, we suppose that  the  eigenenergies are $E_{\pm}^{(n)} = \pm h^{(n)} + \frac{n \omega}{2}$. From Eq. (\ref{cos_h}), we can get that $\cos\left(h^{(0)} T\right) = -\cos\left(h^{(1)} T\right)$. Then, $h^{(1)} T = h^{(0)} T + \pi$, which leads to $h^{(1)} = h^{(0)} + \frac{\omega}{2}$. Finally, $E_+^{(0)} = E_+^{(1)} + \omega$ and $E_-^{(0)} = E_-^{(1)}$. As is well-known, the Floquet quasi-energies are periodic with period $\omega$, and the total quasi-energies should be $E_{\pm}^{(n)} + l\omega$ with $l = 0, \pm1, \pm2, \dots$ Therefore, $E_{\pm}^{(0)}$ and $E_{\pm}^{(1)}$ are equivalent. From now on, we only consider $n=0$ in the $\mathcal{PT}$-symmetric phase.

When $h$ is an imaginary number, $\cos\left(h T\right)>1$, it corresponds to the $\mathcal{PT}$ symmetry spontaneous breaking. There are two kinds of $\mathcal{PT}$ symmetry  broken phases, and their effective Hamiltonians differ by a constant. For simplicity, we assign the right-hand side of Eq. (\ref{cos_h}) to $\Pi(\Delta_0, \Delta_1, \gamma_0, \gamma_1, T_0, T_1)$, namely,
\begin{widetext}
\begin{eqnarray}
\Pi(\Delta_0, \Delta_1, \gamma_0, \gamma_1, T_0, T_1) = \cos \left( h_{1}T_{1}\right) \cos \left(
h_{0}T_{0}\right) +\frac{1 }{4}\left( \gamma _{1}\gamma _{0}-\Delta_{1}
\Delta_{0}\right) T_{0}T_{1}\mathrm{sinc}\left( h_{1}T_{1}\right) \mathrm{sinc}%
\left( h_{0}T_{0}\right).\label{EP}
\end{eqnarray}
\end{widetext}
If $\Pi(\Delta_0, \Delta_1, \gamma_0, \gamma_1, T_0, T_1)$ is greater than $1$, then $n=0$. If $\Pi(\Delta_0, \Delta_1, \gamma_0, \gamma_1, T_0, T_1)$ is less than $-1$, then $n=1$. The exceptional points correspond to $h=0$. From Eq. (\ref{cos_h}), we can easily find that the exceptional points occur when $\Pi(\Delta_0, \Delta_1, \gamma_0, \gamma_1, T_0, T_1) = \pm 1$, where $+$ ($-$) corresponds to $n=0$ ($1$). Unlike the static Hamiltonians $H_j$ whose eigenenergies can only be $0$ in the exceptional points, the quasi-energies of the driven two-level system can be either $0$ for $n=0$ or $\omega/2$ for $n=1$.  Once the parameters $\Delta_j$, $\gamma_j$, $T_j$ of the driving two-level systems are obtained, we can calculate $\Pi(\Delta_0, \Delta_1, \gamma_0, \gamma_1, T_0, T_1)$, from which one can determine whether the $\mathcal{PT}$ symmetry is broken or not.

\subsection{Multiphoton resonance}

For the two-level system with square-wave modulated dissipation and time-independent coupling, the multiphoton resonance refers to the case when the coupling strength $\Delta$ of the two-level system is an integral multiple of the driving frequency $\omega$. A vanishingly small dissipation strength can lead to the $\mathcal{PT}$ symmetry spontaneous breaking in the $(2k-1)$-photon resonance case ($k=1,2\dots$), which has been found in the two-level system with sinusoidal \cite{PhysRevA.92.042103} and square-wave \cite{li2019observation} modulated dissipations.

For the two-level system with a square-wave modulated coupling, one might naively think that the necessary condition for the weak dissipation induced $\mathcal{PT}$ symmetry breaking is that both $\Delta_0$ and $\Delta_1$ are integral multiples of $\omega$. However, it is not the case.
The $\mathcal{PT}$ phase transition induced by the weak dissipation in the multiphoton resonance indicates that $\Pi(\Delta_0, \Delta_1, \gamma_0, \gamma_1, T_0, T_1)$ deviates from $\pm 1$ once the dissipation occurs. We expect that the necessary condition is  $\Pi(\Delta_0, \Delta_1, \gamma_0=0, \gamma_1=0, T_0, T_1)=\pm 1$. From Eq. (\ref{EP}), we can  obtain that
\begin{eqnarray}
&&\Pi(\Delta_0, \Delta_1, \gamma_0=0, \gamma_1=0, T_0, T_1) \nonumber\\
&=& \cos \left( \frac{\Delta_{1}T_{1}}{2}\right) \cos \left(
\frac{\Delta_{0}T_{0}}{2}\right) - \sin \left( \frac{\Delta_{1}T_{1}}{2}\right) \sin%
\left( \frac{\Delta_{0}T_{0}}{2}\right)\nonumber\\
&=& \cos\left( \frac{\Delta_{0}T_{0}+\Delta_{1}T_{1}}{2}\right)\nonumber\\
&=& \cos \left( \frac{\Delta_{\mathrm{eff}} T}{2}\right),\nonumber
\end{eqnarray}
where
\begin{eqnarray}
\Delta_{\mathrm{eff}}=\frac{\Delta_0 T_0 + \Delta_1 T_1}{T}.
\end{eqnarray}
Therefore, the necessary condition for the $\mathcal{PT}$ phase transition induced by the weak dissipation should be $\Delta_{\mathrm{eff}}= k \omega$. In another word, the driving frequency should resonate with the effective coupling strength $\Delta_{\mathrm{eff}}$, rather than $\Delta_0$ or $\Delta_1$. When $k$ is an even number, $\Pi(\Delta_0, \Delta_1, \gamma_0=0, \gamma_1=0, T_0, T_1)=1$. A weak dissipation can lead to $\Pi>1$ which corresponds to  the $\mathcal{PT}$ symmetry broken phase with $n=0$, or $\Pi<1$ which corresponds to  the $\mathcal{PT}$-symmetric phase. Similarly, when $k$ is an odd number, a weak dissipation can lead to $\Pi<-1$ which corresponds to  the $\mathcal{PT}$ symmetry broken phase with $n=1$, or $\Pi>-1$ which corresponds to  the $\mathcal{PT}$-symmetric phase.

\subsubsection{Time-periodic dissipation}

\begin{figure}[htb]
\centering
\includegraphics[width=8cm]{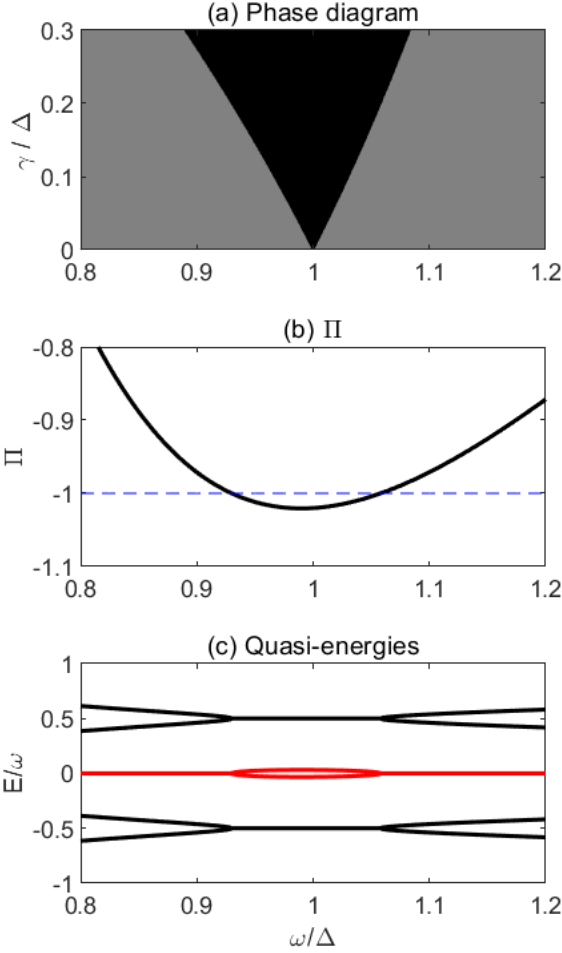}
\caption{ (a) $\mathcal{PT}$ phase diagram near the one-photon resonance, showing $\mathcal{PT}$-symmetric phase (grey), and $\mathcal{PT}$ symmetry broken phases with  $n=1$ (black).
%The blue dash line corresponds to $\gamma/\Delta=0.2$. The triangle, square and circle represent $\omega/\Delta=0.9$, $1$ and $1.1$ respectively.
(b) $\Pi$ (Eq. \ref{EP}) as a function of $\omega/\Delta$ at $\gamma/\Delta=0.2$. The  dash line represents $\Pi=-1$, below which corresponds to $\mathcal{PT}$ symmetry broken phase with $n=1$. (c) Real (black lines) and imaginary (red lines) parts of the quasi-energies as a function of $\omega/\Delta$ at $\gamma/\Delta=0.2$. The other parameters are $\Delta_0=\Delta_1=\Delta=1$,  $T_0=T_1=T/2$, $\gamma_0=\gamma$ and $\gamma_1=0$.}
\label{phase_odd}
\end{figure}
We firstly consider the two-level system with only square-wave modulated dissipation. The coupling strength is time-independent, namely $\Delta_0=\Delta_1=\Delta$, which leads to $\Delta_{\mathrm{eff}}=\Delta$. According to the former analysis, we expect that the $\mathcal{PT}$ phase transition at weak dissipation occurs when $\Delta =k \omega$. However, Li \textit{et al}. only showed the $\mathcal{PT}$-symmetry breaking in $(2k-1)$-photon resonance \cite{li2019observation}, namely $\Delta=(2k-1)\omega$.
In Fig. \ref{phase_odd} (a), we recover the $\mathcal{PT}$ phase diagram near the one-photon resonance in Ref. \cite{li2019observation},  by setting $T_0 = T_1 = T / 2$, $\gamma_0=\gamma$ and $\gamma_1=0$. The boundary of the phase diagram can be determined by either $\Pi(\Delta_0, \Delta_1, \gamma_0, \gamma_1, T_0, T_1)$ (Fig. \ref{phase_odd} (b)), or the imaginary part of the quasi-energies (Fig. \ref{phase_odd} (c)). Near the one-photon resonance region, $\Pi$ is less than $-1$ and the imaginary part of the quasi-energies is nonzero, which indicates that it corresponds to a $\mathcal{PT}$ symmetry broken phase with $n=1$.

\begin{figure}[htb]
\centering
\includegraphics[width=8cm]{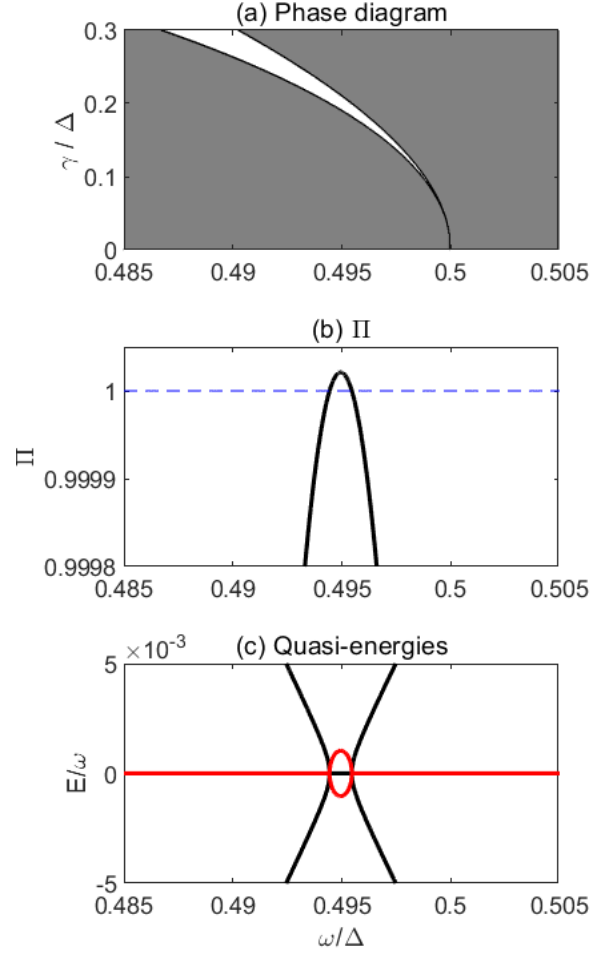}
\caption{(a) $\mathcal{PT}$ phase diagram near the two-photon resonance, showing $\mathcal{PT}$-symmetric phase (grey), and $\mathcal{PT}$ symmetry broken phase with  $n=0$ (white).
%The blue dash line corresponds to $\gamma/\Delta=0.2$. The triangle, square and circle represent $\omega/\Delta=0.488$, $0.495$ and $0.502$ respectively.
(b) $\Pi$ (Eq. \ref{EP}) as a function of $\omega/\Delta$ at $\gamma/\Delta=0.2$. The  dash line represents $\Pi=1$, above which corresponds to $\mathcal{PT}$ symmetry broken phase with $n=0$. (c) Real (black lines) and imaginary (red lines) parts of the quasi-energies as a function of $\omega/\Delta$ at $\gamma/\Delta=0.2$. The other parameters are $\Delta_0=\Delta_1=\Delta=1$, $T_0=T_1=T/2$, $\gamma_0=\gamma$ and $\gamma_1=0$.}
\label{phase_even}
\end{figure}
When we further decrease the driving frequency $\omega$ to the two-photon resonance region, we find that a weak dissipation can also lead to the $\mathcal{PT}$ symmetry breaking, which is not observed in Ref. \cite{li2019observation}. As depicted in Fig. \ref{phase_even} (a), the $\mathcal{PT}$ symmetry broken region is much narrower than that in the one-photon resonance case. Besides, the driving frequency $\omega$ at the phase boundary tends to decrease with increasing $\gamma$. Therefore, the $\mathcal{PT}$ symmetry breaking occurs at the region where $\omega$ is a bit less than $\Delta / 2$. Near the two-photon resonance, $\Pi$ is greater than $1$ and the imaginary part of the quasi-energies is nonzero, which indicates that it corresponds to a $\mathcal{PT}$ symmetry broken phase with $n=0$.

\begin{figure}[htb]
\centering
\includegraphics[width=8cm]{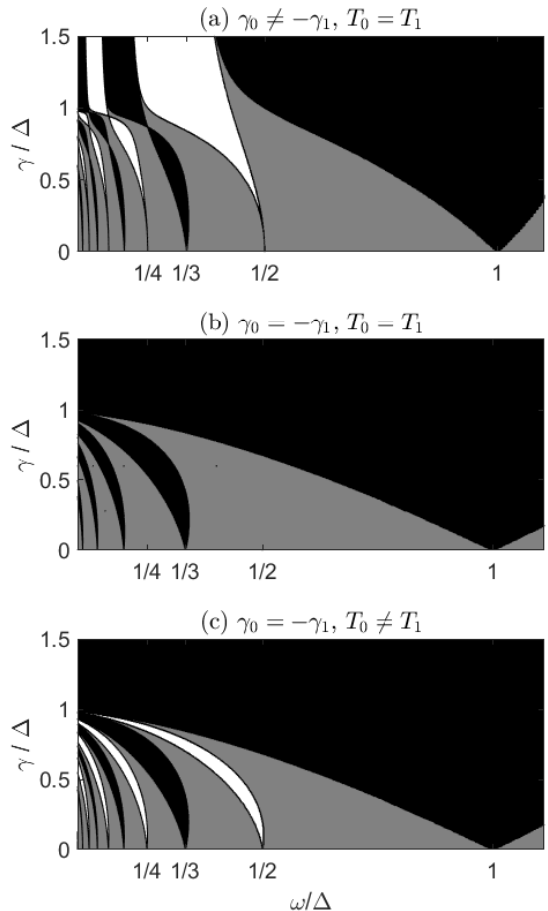}
\caption{$\mathcal{PT}$ phase diagram for time-periodic dissipation near the multiphoton resonance, showing $\mathcal{PT}$-symmetric phase (grey), and $\mathcal{PT}$ symmetry broken phases with  $n=0$ (white) and $n=1$ (black). (a) $\gamma_0=\gamma$, $\gamma_1=0$, $T_0=T_1$. (b) $\gamma_0=-\gamma_1=\gamma$, $T_0=T_1$. (c) $\gamma_0=-\gamma_1=\gamma$, $T_0=0.55T$, $T_1=0.45T$.}
\label{phase_total}
\end{figure}
Fig. \ref{phase_total} (a) is a generalization of Figs. \ref{phase_odd} (a) and \ref{phase_even} (a), which extends the range of $\omega$. The driving two-level system has a much richer phase diagram than the static one. Clearly, a vanishingly small dissipation strength can lead to the $\mathcal{PT}$ symmetry spontaneous breaking in both $(2k-1)$- and $2k$-photon resonances, which is consistent with our criteria $\Delta= k \omega$. To explain the behavior of the $\mathcal{PT}$ symmetry breaking near the $2k$-photon resonance, we reexamine $\Pi(\Delta_0, \Delta_1, \gamma_0, \gamma_1, T_0, T_1)$ in  Eq. (\ref{EP}) in more detail. We suppose that $\gamma_0 = \gamma \ll \Delta$, $\gamma_1 = \lambda \gamma$, $T_0=T_1=T/2$, and $\Delta \simeq 2k \omega$. When $\gamma$ tends to zero, $h_i T_i$ tends to $k \pi$. The first term in the right-hand side of Eq. (\ref{EP}) tends to one while the second term tends to zero. If the second term is greater than zero, it can lead to $\Pi>1$ and the $\mathcal{PT}$ symmetry broken phase with $n=0$. Since $\left( \gamma _{0}\gamma _{1}-\Delta
^{2}\right) T_{0}T_{1}/4$ in the second term is less than zero, one need that $\mathrm{sinc}\left( h_{1}T_{1}\right) \mathrm{sinc}%
\left( h_{0}T_{0}\right)<0$, or $\sin\left( h_{1}T_{1}\right) \sin
\left( h_{0}T_{0}\right)<0$. Then, the condition for the occurrence of $\mathcal{PT}$ symmetry breaking is that one of $h_i T_i$ should be less than $k\pi$, while the other one should be greater than $k\pi$. If $\Delta$ is a bit less than $2k\omega$, a finite $\gamma$ will always decrease $h_i$, which leads to that both $h_i T_i< \Delta T_i / 2 < k\pi$ and $\Pi<1$. Therefore, no $\mathcal{PT}$ symmetry breaking occurs when $\Delta < 2k \omega$. If $\Delta$ is a bit larger than $2k\omega$, one can always find certain $\gamma$ which satisfies the condition for the occurrence of $\mathcal{PT}$ symmetry breaking, as long as $\lambda\ne\pm1$. Fig. \ref{phase_total} (a) corresponds to $\lambda=0$. Therefore,  a finite $\gamma$ can lead to the $\mathcal{PT}$ symmetry breaking near $2k$-photon resonance.

When $\lambda=+1$, namely $\gamma_0=\gamma_1$, the Hamiltonian (\ref{Ht}) becomes time-independent, which is trivial. When $\lambda=-1$, namely $\gamma_0=-\gamma_1=\gamma$, $h_0$ equals to $h_1$. $h_1 T_1 = h_0 T_0$ if $T_0=T_1$, which leads to the $\mathcal{PT}$-symmetric phase with $\Pi<1$ near the $2k$-photon resonance, as shown in Fig. \ref{phase_total} (b). Following the above analysis, we can easily  prove that an imbalanced driving time $T_0 \ne T_1$ can lead to the $\mathcal{PT}$ symmetry breaking when $\gamma_0=-\gamma_1$, as depicted in Fig. \ref{phase_total} (c). The $\mathcal{PT}$ symmetry breaking near $2k$-photon resonance induced by the imbalanced driving time $T_0 \ne T_1$ is more obvious than that induced by $\gamma_0 \ne -\gamma_1$, when the dissipation strength is very weak. Therefore, the imbalanced driving time $T_0 \ne T_1$ is a more efficient method to access the $\mathcal{PT}$ symmetry breaking near $2k$-photon resonance in the experiments.
Figs. \ref{phase_total} (a) and (c) verify our conclusion that the $\mathcal{PT}$ symmetry breaking induced by weak dissipation generally occurs at  both $2k$- and $(2k-1)$-photon resonances, namely $\Delta=k\omega$. The $\mathcal{PT}$ symmetry breaking at $2k$-photon resonance disappears only if $\gamma_0=-\gamma_1$ and $T_0=T_1$, as shown in Fig. \ref{phase_total} (c).

\subsubsection{Time-periodic coupling}

\begin{figure}[htb]
\centering
\includegraphics[width=8cm]{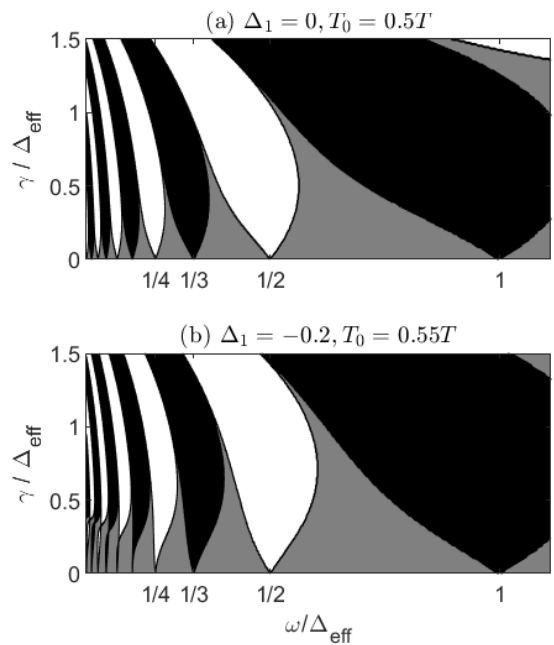}
\caption{$\mathcal{PT}$ phase diagram for time-periodic coupling near the multiphoton resonance, showing $\mathcal{PT}$-symmetric phase (grey), and $\mathcal{PT}$ symmetry broken phases with  $n=0$ (white) and $n=1$ (black). (a) $\Delta_0=1$, $\Delta_1=0$, $T_0=0.5T$. (b) $\Delta_0=1$, $\Delta_1=-0.2$, $T_0=0.55T$.}
\label{phase_coupling}
\end{figure}
For the two-level system with only square-wave modulated coupling,  the dissipation strength is time-independent, namely $\gamma_0=\gamma_1=\gamma$. Fig. \ref{phase_coupling} shows the $\mathcal{PT}$ phase diagram for time-periodic coupling near the multiphoton resonance. Li  \textit{et al}.  studied the influence of the time-periodic coupling on the non-Hermitian two-level system based on a simpler model with $\Delta_0=\Delta$, $\Delta_1=0$ and $T_0=T_1=T/2$ \cite{li2019observation}, which corresponds to Fig. \ref{phase_coupling} (a). They concluded that the $\mathcal{PT}$ phase transition induced by the weak dissipation occurs at $\Delta = 2k \omega$, which is consistent with our results $\Delta_{\mathrm{eff}}=k \omega$ due to $\Delta_{\mathrm{eff}}=\Delta/2$. Fig. \ref{phase_coupling} (b) introduces a nonzero $\Delta_1$ and imbalanced driving time $T_0 \ne T_1$, which cannot be explained by Ref. \cite{li2019observation}. However, $\Delta_{\mathrm{eff}}=k \omega$ can still provide the right condition at which the $\mathcal{PT}$ phase transitions occur. The $\mathcal{PT}$ symmetry broken phase with $n=0 $ ($1$) occurs when $k$ is even (odd), which is also consistent with our former analysis.

\subsection{High frequency limit: $T \rightarrow 0$}

\begin{figure}[htb]
\centering
\includegraphics[width=8cm]{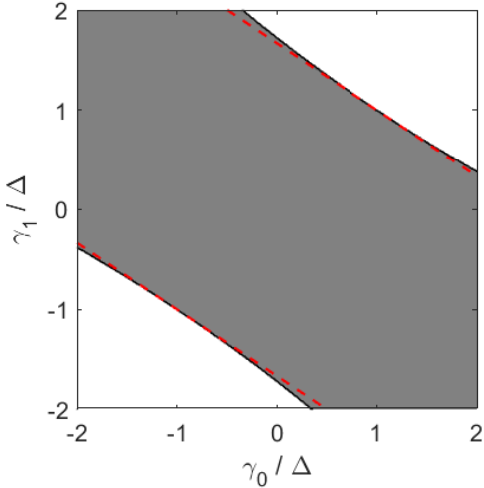}
\caption{ $\mathcal{PT}$ phase diagram, showing $\mathcal{PT}$-symmetric phase (grey), and $\mathcal{PT}$ symmetry broken phase with  $n=0$ (white)  at $\Delta_0=\Delta_1=\Delta=1$, $\omega=3$, $T_0=0.4T$, $T_1=0.6T$. The red dash lines refer to the analytical results in the high frequency limit.}
\label{small_T}
\end{figure}
If the driving frequency is very large, namely $\omega \gg \Delta_j, \gamma_j$, the period $T$ tends to zero. We suppose that $T_0$ and $T_1$ are of same order as $T$. Expanding $\Pi(\Delta_0, \Delta_1, \gamma_0, \gamma_1, T_0, T_1)$ to the second order of $T$, we obtain
%\begin{widetext}
\begin{eqnarray}
&&\Pi(\Delta_0, \Delta_1, \gamma_0, \gamma_1, T_0, T_1) \nonumber\\
&\simeq& \left(1 - \frac{h_{1}^2T_{1}^2}{2}\right)  \left(
1 - \frac{h_{0}^2T_{0}^2}{2}\right) +\frac{1 }{4}\left( \gamma _{1}\gamma _{0}-\Delta_{1} \Delta_{0} \right) T_{1}T_{0} \nonumber\\
&\simeq& 1 + \frac{1 }{4}\left[\left( \gamma _{1}\gamma _{0}-\Delta_{1} \Delta_{0} \right) T_{1}T_{0} - 2 h_{0}^2T_{0}^2 - 2 h_{1}^2T_{1}^2 \right]\nonumber\\
&=& 1 + \frac{1}{8} \left[ \left(\gamma_0 T_0 + \gamma_1 T_1\right)^2 - \left(\Delta_0 T_0 + \Delta_1 T_1\right)^2 \right]\nonumber\\
&=& 1 + \frac{1}{8} \left( \gamma_{\mathrm{eff}}^2 - \Delta_{\mathrm{eff}}^2 \right) T^2,
\end{eqnarray}
%\end{widetext}
where
\begin{eqnarray}
\gamma_{\mathrm{eff}}=\frac{\gamma_0 T_0 + \gamma_1 T_1}{T}.
\end{eqnarray}
Therefore, the exceptional points, as well as the $\mathcal{PT}$ phase boundary, are located at $\gamma_{\mathrm{eff}}=\pm \Delta_{\mathrm{eff}}$.
If $|\gamma_{\mathrm{eff}}|<|\Delta_{\mathrm{eff}}|$, it corresponds to the $\mathcal{PT}$-symmetric phase. Otherwise, the $\mathcal{PT}$ symmetry is broken with $n=0$. Alternately, if we expand Eqs. (\ref{J})-(\ref{Gamma_z}) to the lowest order of $T$, we find that
\begin{eqnarray}
J \simeq  \Delta_{\mathrm{eff}},\qquad
\Gamma_y \simeq 0, \qquad
\Gamma_z \simeq \gamma_{\mathrm{eff}},
\end{eqnarray}
which give rise to the following effective Hamiltonian,
\begin{eqnarray}
H_{\mathrm{eff}} \simeq \frac{\Delta_{\mathrm{eff}} }{2}\sigma _{x}+\mathrm{i}\frac{\gamma_{\mathrm{eff}} }{2}\sigma _{z}.
\end{eqnarray}
It leads to  {the} same $\mathcal{PT}$ phase boundary. In a word, we find that when the driving frequency is very large, the Floquet effective Hamiltonian is equivalent to a static one with time-averaged coupling and dissipation strength. When $\Delta_0 \simeq -\Delta_1$, $\Delta_{\mathrm{eff}}$ tends to zero and one can easily achieve the $\mathcal{PT}$ symmetry broken phase no matter how large $\Delta_j$ is.  When $\gamma_0 \simeq -\gamma_1$, $\gamma_{\mathrm{eff}}$ tends to zero and one can easily preserve the $\mathcal{PT}$ symmetry no matter how large $\gamma_j$ is.

Fig. \ref{small_T} shows the $\mathcal{PT}$ phase diagram at $\Delta_0=\Delta_1=\Delta$, $\omega/\Delta=3$ and $T_0/T_1=2/3$. The phase boundary $\gamma_{\mathrm{eff}}=\pm \Delta_{\mathrm{eff}}$ fits well with the exact results.

\section{Conclusions}

We study a non-Hermitian two-level system with square-wave modulated dissipation and coupling. Two time-independent Hamiltonians $H_0$ and $H_1$ appear alternately.   {Comparing with the formerly well-known sinusoidal modulation, the square-wave modulation has three advantages: Firstly,  exact analytical solutions can be achieved by employing the Floquet theory. Secondly, the $\mathcal{PT}$ phase diagram becomes  richer.  Thirdly, the square-wave modulation has been realized in the ultracold atoms experiment \cite{li2019observation}.}

Based on the Floquet theory, we achieve an effective  Hamiltonian with $\mathcal{PT}$ symmetry.
We define a parameter $\Pi(\Delta_0, \Delta_1, \gamma_0, \gamma_1, T_0, T_1)$, from which one can derive the boundaries of the $\mathcal{PT}$ phase diagram exactly.  The driving two-level system has
a much richer phase diagram than the static one. Two kinds of $\mathcal{PT}$ symmetry broken phases are found whose effective Hamiltonians differ by a constant $\omega / 2$. When $\Pi>1$, the $\mathcal{PT}$ symmetry broken phase with $n=0$ occurs. When $\Pi<-1$, the $\mathcal{PT}$ symmetry broken phase with $n=1$ occurs. When $-1<\Pi<1$, the $\mathcal{PT}$ symmetry is preserved.

With the help of $\Pi$, we firstly study the $\mathcal{PT}$ phase transition with only square-wave modulated dissipation near multiphoton resonance. The coupling strength is time-independent with $\Delta_0=\Delta_1=\Delta$. A weak dissipation can lead to the $\mathcal{PT}$ symmetry  breaking near the $(2k-1)$-photon resonance ($\Delta=(2k-1)\omega$), which  has been observed in the ultracold atoms experiment \cite{li2019observation}. We predict that the $\mathcal{PT}$ symmetry  breaking near the $2k$-photon resonance ($\Delta=2k\omega$), can also happen  as long as the dissipation strengths or the driving times are imbalanced, with $\gamma_0 \ne -\gamma_1$ or $T_0 \ne T_1$. Our studies pave a way to access the $\mathcal{PT}$ symmetry broken phase near the $2k$-photon resonance in the experiments. For the $\mathcal{PT}$ phase transition with  square-wave modulated coupling, we define an effective coupling strength $\Delta_{\mathrm{eff}}=\left(\Delta_0 T_0 + \Delta_1 T_1\right)/T$. The weak dissipation induced $\mathcal{PT}$ symmetry  breaking can occur only if $\Delta_{\mathrm{eff}}=k\omega$.

In the high frequency limit, we achieve a simple relation $\gamma_{\mathrm{eff}} = \pm \Delta_{\mathrm{eff}}$, which gives the $\mathcal{PT}$ phase boundary. When $\Delta_0 \simeq -\Delta_1$, one can easily achieve the $\mathcal{PT}$ symmetry broken phase no matter how large the coupling strength $|\Delta_j|$ is.  When $\gamma_0 \simeq -\gamma_1$, one can easily preserve the $\mathcal{PT}$ symmetry no matter how large the dissipation strength $|\gamma_j|$ is.

\section*{ACKNOWLEDGEMENTS}
This work is supported by the National Science
Foundation of China (Grant Nos. 11674285 and  11834005), the National Key
Research and Development Program of China (Grant No. 2017YFA0303002).

%\section*{References}

\end{document}